\documentclass[aps,prb,floatfix,twocolumn,superscriptaddress]{revtex4-1}
\usepackage{graphicx}
\usepackage[english]{babel}
\usepackage{amsmath}
\usepackage{amssymb}
\usepackage{tensor}

\newcommand{\vk}{\mathbf{k}}

\newcommand{\vR}{\mathbf{R}}
\newcommand{\be}{\begin{eqnarray}}
\newcommand{\ee}{\end{eqnarray}}
\newcommand{\p}{\partial}

\def\ket#1{|#1\rangle}
\def\bra#1{\langle #1 |}
\def\ep#1{\langle #1 \rangle}

\begin{document}

\title{Comparison of finite-temperature topological indicators based on Uhlmann connection}

\author{Ye Zhang}
\affiliation{College of Physics, Sichuan University, Chengdu, Sichuan 610064, China}
\email{heyan$_$ctp@scu.edu.cn}

\author{Aixin Pi}
\affiliation{College of Physics, Sichuan University, Chengdu, Sichuan 610064, China}

\author{Yan He}
\affiliation{College of Physics, Sichuan University, Chengdu, Sichuan 610064, China}
\email{heyan$_$ctp@scu.edu.cn}

\author{Chih-Chun Chien}
\affiliation{School of Natural Sciences, University of California, Merced, CA 95343, USA.}
\email{cchien5@ucmerced.edu}

\begin{abstract}
Two indicators of finite-temperature topological properties based on the Uhlmann connection, one generalizing the Wilson loop to the Uhlmann-Wilson loop and the other generalizing the Berry phase to the Uhlmann phase, are constructed explicitly for a time-reversal invariant topological insulators with a $Z_2$ index. While the phases of the eigenvalues of the Wilson loop reflect the $Z_2$ index of the model at zero temperature, it is found that the signature from the Uhlmann-Wilson loop gradually fades away as temperature increases. On the other hand, the Berry phase exhibits quantization due to the underlying holonomy group. The Uhlmann phase retains the quantization at finite temperatures and serves as an indicator of topological properties. A phase diagram showing where jumps of the Uhlmann phase can be found is presented. By modifying the model to allow higher winding numbers, finite-temperature topological regimes sandwiched between trivial regimes at high and low temperatures may emerge.
\end{abstract}

\maketitle

\section{Introduction}
In recent years, there has been a huge progress in understanding the topological properties of quantum matter, such as topological insulators and topological superconductors \cite{Zhang_TIRev,Kane_TIRev,Chiu2016,Asboth2016,Bernevig_book}. One important achievement is the ten-fold way classification \cite{Chiu2016} of free fermion systems according to three types of discrete symmetries. Other than the symmetry class A, almost all topological non-trivial phases are protected by certain symmetries, leading to a more general concept of symmetry protected topological phases. Despite the rapid developments over the past decade, most works focus on the topology of the ground state at zero temperature. When a quantum system is at finite temperature or out of equilibrium, one has to consider the topological properties of a mixed state that represents a statistical ensemble. The research on the topology of mixed states is an active field~\cite{Huang14,Viyuela14,Diehl,Asorey19,HouPRB20,Unanyan20} and is the main focus of this work.

The basis of the ground state topology is usually built on the Berry connection \cite{Berry}, which gives a geometric phase to the wavefunction under cyclic adiabatic evolution. From the Berry connection, one can obtain the Berry phase, Berry curvature, and other topological characteristics. In parallel, several pioneer works \cite{Sjoqvist00,Viyuela14,Diehl,Diehl18} tried to generalize the concept of geometric connection of pure states to mixed states. Among these attempts, the Uhlmann connection \cite{Uhlmann,Uhlmann1,Uhlmann2,Mera17} is a promising notion defined on the fiber bundle from full-rank density matrices. The key point of the Uhlmann connection is the parallel condition between the amplitudes of density matrices, which will be briefly reviewed later. The Uhlmann connection and its associated quantities have been applied to understand the topology of several one-dimensional or two-dimensional models \cite{Viyuela14,Viyuela15,Viyuela2,Huang14}, spin systems~\cite{Galindo21,HouPRA21}, and others~\cite{Villavicencio21}. It has been found that at certain critical temperature, there exists a transition from a topologically non-trivial phase to a trivial phase accompanied by a jump in the Uhlmann phase. The topology change comes from the Uhlmann holonomy, as explained in Refs.~\cite{Viyuela15,HouPRA21}.

Here we will apply the Uhlmann connection to investigate an exemplary time-reversal invariant topological insulators (TI) at finite temperatures. The time-reversal invariant TI, or quantum spin Hall effect, has been proposed by Kane and Mele for graphene \cite{Kane-Z2} but was not successfully realized due to the weak spin-orbital coupling. The non-trivial phase was also proposed to be realizable in the HgTe quantum well \cite{BHZ-1}, which was experimentally observed \cite{BHZ-2}. The band structure of the HgTe quantum well can be captured by a simple 4-band model, which is also known as Bernevig-Hughes-Zhang (BHZ) model. This model will be the main platform here for testing finite-temperature topological indicators based on the Uhlmann connection. The $Z_2$ topological index characterizing the topology of the time-reversal invariant TI at zero temperature is more subtle than the ordinary Chern number. For example, the calculation of the Fu-Kane invariant \cite{Fu-Z2} requires the use of globally defined eigen-functions. This drawback prompted later works \cite{Yu-Z2} to make use of gauge invariant quantities, such as the Wilson loop, to indicate the underlying topology. Due to the strong analogy between the Berry and Uhlmann connections, we propose to implement both the Uhlmann-Wilson loop and the Uhlmann phase as indicators to study the finite-temperature topology, using the BHZ model as an example. We mention that the BHZ model has been studied by using the Uhlmann connection in Ref.~\cite{Huang14}, but the detailed analysis and the results are different.

While the Uhlmann-Wilson loop generalizes the Wilson loop that gives the $Z_2$ index at zero temperature, we found that the finite-temperature contributions gradually reduce the magnitude of the phases of it eigenvalues. As a consequence, the difference between the topological and trivial cases becomes less prominent as temperature increases. In contrast, the Uhlmann phase generalizes the Berry phase and still reflects the holonomy group. For the BHZ model, the holonomy group is the $Z_2$ group and the Uhlmann phase remains quantized at finite temperatures. Therefore, the Uhlmann phase can still clearly distinguish the topological regime from the trivial one. Moreover, a finite-temperature topological regime sandwiched by topologically trivial regimes at both lower and higher temperatures may emerge if the model allows higher winding numbers. Such a possibility will be demonstrated by a modification of the BHZ model. Moreover, it will be pointed out that the $Z_2$ Uhlmann holonomy group is not directly associated with the $Z_2$ index of the ground state although their topological regimes agree at zero temperature. Such a subtlety shows the rich topological properties and their indicators at finite temperatures.

The rest of the paper is organized as follows. In Section \ref{sec:review}, we briefly review the Uhlmann process from a geometric point of view. The properties of the BHZ model will also be reviewed. In Section \ref{sec:Wilson}, we present the topological properties of the BHZ model according to the Wilson loop and its Uhlmann-Wilson generalization. The gauge-invariant $Z_2$ index of the BHZ model at $T=0$ is shown to lose its signature as temperature increases. In Section \ref{sec:Uphase}, the Uhlmann phase is introduced as a generalization of the Berry phase. The quantized Uhlmann phase allows a clear distinction of finite-temperature topological regimes. We offer semi-analytical explanations of the behavior from both approaches. Finally, 
we conclude our study in Section \ref{sec:Con}. The Appendix summarizes some details and properties mentioned in the main text.

\section{Brief review of concept and model}\label{sec:review}
\subsection{Uhlmann process}
\label{sec-U}
We begin by briefly reviewing the Uhlmann process, which is a finite-temperature generalization of the Berry process at zero temperatures. The Uhlmann process is based on the concept of the Uhlmann 
connection, an analogue of the Berry connection. Before going into details, we first present the Berry connection in a more geometric point of view. The Berry connection is defined for a given eigenstate $\ket{\psi(r)}=e^{i\theta(r)}\ket{u(r)}$ with some parameter $r$. Due to the arbitrary phase $\theta(r)$, $\ket{\psi(r)}$ forms a $U(1)$ fiber bundle over the parameter space. Two different states are said to be parallel to each other if \cite{GPhase_book} $\ep{\psi(r_1)|\psi(r_2)}>0$. The infinitesimal version of the parallel condition can be written as $\ep{\psi(r)|\p_r|\psi(r)}=0$, which gives rise to the Berry connection
\be
A_r=\p_r\theta=-i\ep{u|\p_r|u}.
\ee

At finite temperature, the density matrix $\rho$ of a mixed state should be used in place of the wave function. The spectral decomposition of the density matrix gives $\rho=\sum_i p_i\ket{u_i}\bra{u_i}$ with the eigenstates $\ket{u_i}$. In thermal equilibrium, the weight $p_n$ is proportional to the Boltzmann factor and all the eigenstates contribute to the density matrix. At $T=0$, the density matrix reduces to a projection operator of a pure state. The amplitude decomposition of the density matrix is given by
\be
\rho=ww^{\dagger},\qquad w=\sqrt{\rho}\,U.
\ee
Here $w$ may be thought of as the counterpart of the wave function for a mixed state. However, $w$ is not uniquely determined for a given $\rho$. Just as a pure state can have an arbitrary U$(1)$ phase, the definition of $w$ also includes an arbitrary unitary matrix $U$. The amplitudes actually forms a Hilbert space like the wave functions do. In this space, one can introduce the Hilbert-Schmidt inner product \cite{GPhase_book} $(w_1,w_2)\equiv\textrm{Tr}(w_1^\dag w_2)$.

The crucial idea of the connection in a fiber bundle is the parallel condition. Analogous to the case of pure states, one may attempt to define a parallel condition for a pair of amplitudes as $(w_1,w_2)>0$. Nevertheless, Uhlmann \cite{Uhlmann} proposed a more stringent parallel condition:
\be
w_1^{\dagger}w_2=w_2^{\dagger}w_1=X>0.
\ee
Here $X>0$ means $X$ is a Hermitian and positive definite matrix.
Given two different amplitudes $w_1=\sqrt{\rho_1}U_1$ and $w_2=\sqrt{\rho_2}U_2$, the above parallel condition leads to a relation between $U_1$ and $U_2$. Note that
\be
X^2=w_1^{\dagger}w_2w_2^{\dagger}w_1=U_1^{\dagger}\sqrt{\rho_1}\rho_2\sqrt{\rho_1}U_1
\ee
implies
\be
X=U_1^{\dagger}\sqrt{\sqrt{\rho_1}\rho_2\sqrt{\rho_1}}\,U_1.
\ee
Combining the equation with the parallel condition, we find the phase factor of $w_2$ relative to $w_1$ as
\be
U_{21}\equiv U_2U_1^{\dagger}=\sqrt{\rho_2^{-1}}\sqrt{\rho_1^{-1}}\sqrt{\sqrt{\rho_1}\rho_2\sqrt{\rho_1}}.
\label{U21}
\ee
This may be thought of as a finite version of the Uhlmann connection between $w_2$ and $w_1$. We remark that the derivation assumes all the $\rho$'s are full-rank matrices. Moreover, the parallel condition is not transitive, which is a general feature of curved space. Therefore, the relative phase factor is path dependent. For example, if one considers three different states $w_1$, $w_2$, and $w_3$. The relative phase factor between $w_3$ and $w_1$ can be computed in two ways. First, one can go from $w_1$ to $w_2$ and then to $w_3$. The phase factor is $U_{32}U_{21}$ in this case. Second, one can directly go from $w_1$ to $w_3$, which generate a phase factor $U_{31}$. From Eq.~(\ref{U21}), $U_{31}\neq U_{32}U_{21}$ in general.
In practice, Eq.~(\ref{U21}) may be used to compute the phase factor between states with small parameter differences.

Sometimes, it is more convenient to work with an infinitesimal version of the Uhlmann connection. We consider a pair of density matrices with a small parameter difference: $\rho_1=\rho$ and $\rho_2=\rho+\Delta k_\mu\p_\mu\rho$. Here we assume that $\rho$ depends on the parameter $k_\mu$. The difference in the parameter is $\Delta k_\mu$, and $\p_\mu=\frac{\p}{\p k_\mu}$ for abbreviation. After some algebra, the infinitesimal Uhlmann connection is given by
\be
 A^U_\mu=\p_\mu U U^{\dagger}.
\ee
Note that $A^U_\mu$ defined above is anti-Hermitian. By using the spectral expansion of the density matrix, an explicit expression of the Uhlmann connection is shown to be
\be
A^U_\mu=\sum_{i,j}\ket{u_i}\bra{u_i}\frac{[\p_\mu\sqrt{\rho},\,\sqrt{\rho}]}{p_i+p_j}\ket{u_j}\bra{u_j}.
\label{AU}
\ee
Meanwhile, the following expression will also be useful:
\be
A^U_{\mu}=\sum_{i,j}\frac{(\sqrt{p_i}-\sqrt{p_j})^2}{p_i+p_j}\ket{u_i}\bra{u_i}\p_\mu\ket{u_j}\bra{u_j}.
\label{AU1}
\ee
The derivation is in Appendix~\ref{app:AU}.

We remark that the Uhlmann connection is also a $U(n)$ non-Abelian gauge field over a parameter space. Here $n$ may be associated with the number of bands. Some features of the Uhlmannn connection are mentioned here. Since the definition requires $\rho$ to be non-singular without any zero eigenvalue, Eq.~(\ref{AU}) cannot be applied to pure states. However, it has been shown \cite{Viyuela14} that in certain cases, the Uhlmann phase obtained from $A_U$ approaches the Berry phase as $T\to 0$. A serious drawback of the Uhlmann connection is that $A_U$ is always non-singular and supports a global section of the underlying $U(n)$ bundle. This implies that the the $U(n)$ bundle is topologically trivial and all characteristic classes, such as the Chern class and Chern character, computed from the Uhlmann curvature will vanish \cite{Diehl}. To overcome this difficulty, Ref. \cite{Viyuela2} proposes the Uhlmann number, which approaches the Chern number as $T\to 0$. Ref.~\cite{HeChern18} also suggests a modified Chern-number formula to extract non-vanishing results from the Uhlmann connection.

\subsection{Four-band Bernevig-Hughes-Zhang model}
\label{sec-BHZ}
We also review a prototype of time-reversal invariant topological insulators, the four-band Bernevig-Hughes-Zhang (BHZ) model. Before we consider this model at finite temperatures, we first describe its topological character at zero temperature. The Hamiltonian of BHZ model is given by
\be
H=\left(
    \begin{array}{cc}
      H_0(\vk) &  H_1 \\
      H_1^{\dag} & H_0^*(-\vk)
    \end{array}
  \right).
\ee
The corresponding wave function is $\psi=(\psi_{1\uparrow},\psi_{2\uparrow},\psi_{1\downarrow},\psi_{2\downarrow})^T$, where the index $i=1,2$ labels the two orbitals and the up or down arrow labels the spin. $H_0$ is the Qi-Wu-Zhang model \cite{Qi1} given by
\be
H_0=\sin k_x \sigma_1+\sin k_y \sigma_2+(m+\cos k_x+\cos k_y)\sigma_3.
\label{QWZ}
\ee
Here $\sigma_i$ for $i=1,2,3$ are the Pauli matrices. The model of Eq.~(\ref{QWZ}) is an ordinary Chern insulator with the Chern number
\be
C=\left\{
    \begin{array}{ll}
      1, & 0<m<2;\\
      -1, & -2<m<0; \\
      0, & |m|>2.
    \end{array}
  \right.
\ee
The $H_1$ term is given by
\be
H_1=\left(
    \begin{array}{cc}
      0 &  \gamma \\
      -\gamma & 0
    \end{array}
  \right),
\ee
which is included to break the $S_z$ conservation and the inversion symmetry. The corresponding time-reversal (TR) operator is $U_TK$ with $U_T=i\sigma_2$ and $K$ denoting the complex conjugation operator. The topology of the BHZ model is protected by time-reversal symmetry because of the TR invariant condition
\be
U_T^\dag H^*(\vk) U_T=H(-\vk).
\ee
Due to the TR symmetry condition, the lowest two bands are degenerate at the four time-reversal invariant momentum points $\vk_1=0$, $\vk_2=(\pm\pi,0)$, $\vk_3=(0,\pm\pi)$, and $\vk_4=(\pm\pi,\pm\pi)$. Thus, one cannot define the Chern number for these two bands separately. However, the total Chern number of those two band is identically zero because they have opposite Chern numbers due to the TR symmetry.

Although the total Chern number is always zero, the non-trivial topology can be characterized by the $Z_2$ index. One way to compute the $Z_2$ index is through the Fu-Kane invariant \cite{Fu-Z2} summarized in Appendix~\ref{app:FK}.
The Fu-Kane index has the advantage of being computationally manageable since it only involves an evaluation of the matrix $\bar{W}$ shown in Eq.~\eqref{eq:W} at the four TR invariant momentum points. However, the evaluation of the matrix $\bar{W}$ requires the use of a globally defined eigenstate, which exists in principle but is difficult to find. 
For the BHZ model with small $\gamma$, the Fu-Kane invariant is given by
\be
I_{FK}=\left\{
         \begin{array}{ll}
           -1, & |m|<2, \\
           1, & |m|>2.
         \end{array}
       \right.
\ee
Therefore, the condition of non-trivial topology of the BHZ model is almost the same as that of the QWZ model. At $T=0$, the emergence of edge states in a system with open boundary condition may be considered as another topological property~\cite{Zhang_TIRev,Kane_TIRev,Chiu2016}. The edge states of the BHZ model are reviewed in Appendix~\ref{app:FK}.

\section{Topology according to Wilson loop}\label{sec:Wilson}
\subsection{Topology at zero temperature}
Instead of the Fu-Kane index, there are other works proposing manifestly gauge invariant methods to compute the $Z_2$ index. Here we follow the method based on the Wilson loop or Wannier center \cite{Yu-Z2} by defining a Wilson line operator across a given link on a lattice in momentum space. The matrix element is given by
\be
W^{mn}_{i,i+1}(k_y)=\ep{u_m(k_{x,i},k_y)|u_n(k_{x,i+1},k_y)}.
\ee
Here $\ket{u_m}$ denotes the eigenstate in momentum space and the indices $m,n$ run through all the occupied bands. In the case of the half-filled BHZ model, $W_{i,i+1}$ is a 2 by 2 matrix. The Wilson loop with fixed $k_y$ can be obtained from
\be
W(k_y)=W_{0,1}W_{1,2}W_{2,3}\cdots W_{N-1,N}W_{N,0}.
\label{W-k}
\ee
Here $N$ is the lattice number along the $x$-axis. We note the Wilson line depends on the gauge choice of the eigenstates. Under the transformation $\ket{u(\vk)}\to\ket{u_n(\vk)}e^{i\theta(\vk)}$, we find that $W_{i,i+1}\to W_{i,i+1}e^{i\theta(k_{x,i+1},k_y)-i\theta(k_{x,i},k_y)}$. If a closed loop is traversed in Eq.~(\ref{W-k}), all the arbitrary gauge dependence cancels out and the Wilson loop is manifestly gauge invariant. In the continuum limit, the above Wilson line can also be expressed in terms of the non-Abelian Berry connection as
\be
W_{i,i+1}(k_y)&=&\mathcal{P}\exp\Big(i \int_{k_{x,i}}^{k_{x,i+1}}dk' A_x(k',k_y)\Big) \nonumber \\
&&\approx\exp\Big(i A_x(k_{x,i},k_y)\Delta k\Big),\\
A^{mn}_{\mu}(\vk)&=&-i\ep{u_m(\vk)|\frac{\p}{\p k_{\mu}}|u_n(\vk)}.
\ee
Here $\mu=x,y$, $\Delta k=k_{x,i+1}-k_{x,i}$, and $\mathcal{P}$ denotes the path ordering of the following integral. Therefore, the Wilson loop can be written as
\be
W(k_y)=\mathcal{P}\exp\Big(i\oint_C A_{\mu}(\vk)d k_{\mu}\Big).
\label{W-ky}
\ee
The integral contour $C$ is the loop with fixed $k_y$ while $k_x$ varies from $0$ to $2\pi$.
According to the Stoke theorem, the line integral of the Berry connection along a close loop equals to the surface integral of the Berry curvature, which also demonstrates that the Wilson loop is gauge invariant.

With the Wilson loop in hand, we define the $Z_2$ index by the phase of the eigenvalues of $W(k_y)$. Since $W(k_y)$ is a unitary matrix, its eigenvalues $\lambda_n(k_y)$ are unit-modulus complex number. For given $k_y$, we introduce
\be
\theta(k_y)=\arg [\lambda_n(k_y)].
\ee
Here arg denotes the phase angle (or argument) of a complex number. For the BHZ model, there are only two arguments $\theta_{1,2}(k_y)$. Since det$W=1$, we always have $\theta_1=-\theta_2$. At the TR invariant momentum point $k_y=0$ or $k_y=\pi$, the Wilson loop $W$ has degenerate eigenvalues due to the TR symmetry. Therefore, we find  $\theta_1(0)=\theta_2(0)$ and $\theta_1(\pi)=\theta_2(\pi)$.

\begin{figure}
\centering
\includegraphics[width=\columnwidth]{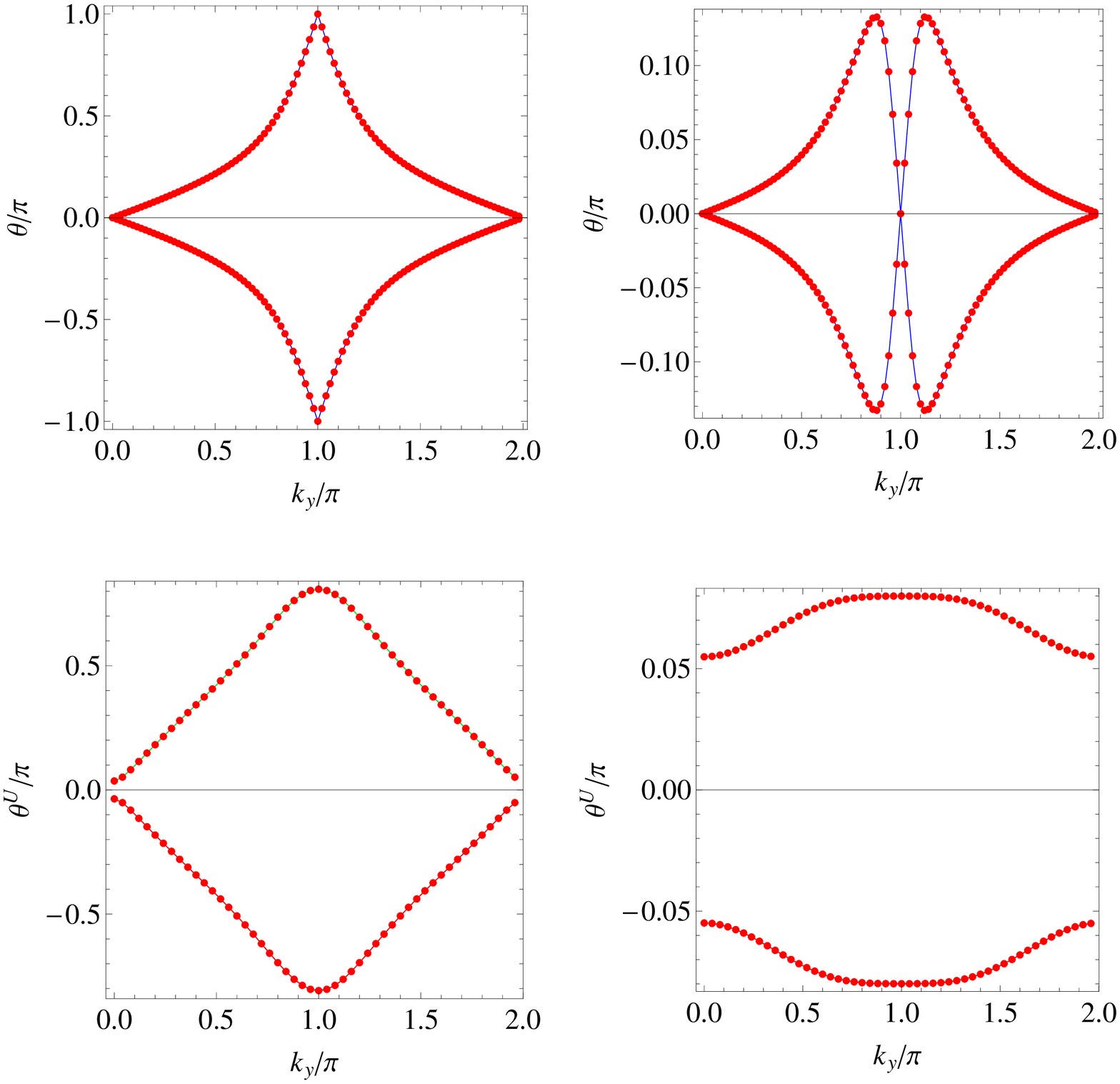}
\caption{Top row: $\theta$ of the eigenvalues of the Wilson loop $W(k_y)$ of Eq. (\ref{W-k}) for the BHZ model as a function of $k_y$ with $m=0.8$ and $\gamma=0.2$ (left) and $m=2.2$ and $\gamma=0.2$ (right). Bottom row: $\theta^U$ of the eigenvalue of the Uhlmann Wilson loop $V(k_y)$ of the BHZ model as a function of $k_y$ with $m=0.8$, $\gamma=0.2$, and $T=0.4$ (left) and $T=2.4$ (right).}
\label{W-loop}
\end{figure}

To understand the topology  from a different perspective, we plot $\theta_1=-\theta_2$ as a function of $k_y$ that shows two different types of behavior in the top row of Figure \ref{W-loop}. In the left panel, we assume $m=0.8$ and $\gamma=0.2$ corresponding to the topological case. The two phases start from zero at $k_y=0$ and gradually increases to $\theta_1=\pi$ and $\theta_2=-\pi$ at $k_y=\pi$. Note that $\pm\pi$ are the same modulo $2\pi$, thus we have $\theta_1=\theta_2$ at $k_y=0,\pi$ as required by the TR symmetry. As $k_y$ further increases to $2\pi$, $\theta_{1,2}$ come back to zero. In this case, one can see that the phase angle $\theta_{1,2}$ have traveled around a full circle, which signals the non-trivial topology. On the other hand, for the case with $m=2.2$ and $\gamma=0.2$ in the right panel of Figure \ref{W-loop}, we find that $\theta_{1,2}$ depart from zero not too far before coming back to zero again. The phases never make a full circle in the latter case, and this represents the trivial topology.

\subsection{Topology at finite temperature}
\label{sec-topo}
We now study finite-temperature topological properties of the BHZ model. Away from zero temperature, the Fermi distribution deviates from the step function, and all the bands have non-vanishing occupation. It is not possible to concentrate only on the occupied bands and define the non-Abelian Berry connection as the zero-temperature case. Since all the bands contribute at finite temperatures, we will use the Uhlmann connection to replace its zero-temperature counterpart and explore the topology at finite temperatures.

Following the idea, the Berry connection in the Wilson loop is replaced by the Uhlmann connection. The result is the Uhlmann-Wilson loop along a closed curve $C$ in the parameter space given by
\be
V=\mathcal{P}\exp\Big(\oint_C A^U_{\mu}d k_{\mu}\Big).
\label{V-loop}
\ee
Again, the integral is under the path ordering. For the BHZ model, the Uhlmann Wilson loop is a 4 by 4 matrix. At zero temperature, we have seen that the topology is reflected by the phase of the eigenvalues of the Wilson loop. As a finite $T$ counterpart, we define the following phase angle for a fixed $k_y$:
\be
\theta^U_n(k_y)=\arg [\lambda_n(k_y)].
\ee
Here $\lambda_n$ is the $n$-th eigenvalue of the Uhlmann-Wilson loop $V$.

To compute the Uhlmann-Wilson loop, we discretize the momentum space by a lattice and calculate the thermal-equilibrium density matrix $\rho(\vk)$ at each site. For a given $k_y$, the Uhlmann-Wilson line across a link on the momentum-space lattice can be obtained by
\be
V_{i,i+1}(k_y)=\sqrt{\rho_2^{-1}}\sqrt{\rho_1^{-1}}\sqrt{\sqrt{\rho_1}\rho_2\sqrt{\rho_1}}.
\ee
Here $\rho_1=\rho(k_{x,i},k_y)$ and $\rho_2=\rho(k_{x,i+1},k_y)$.
Finally, the Uhlmann Wilson loop is given by the product around a loop:
\be
V(k_y)=V_{0,1}V_{1,2}V_{2,3}\cdots V_{N-1,N}V_{N,0}.
\ee

The numerical results of $\theta_n$ from the Uhlmann-Wilson loop is shown in the bottom row of Figure \ref{W-loop} for selected values of temperature. An interesting feature is the relation
\be
\theta_1^U=\theta_2^U=-\theta_3^U=-\theta_4^U\label{th-U}.
\ee
A qualitative understanding of this feature is given in Appendix~\ref{app:UW}. Although the phase $\theta_n^U$ approaches $\theta$ from the Wilson loop at zero temperature as $T\rightarrow 0$, we observe some important differences at finite temperature. Firstly, $\theta^U$'s are no longer pinned to zero at the TR invariant momentum points away from $T=0$. At low temperatures, when the system traverses a loop from $k_y=0$ to $2\pi$, $\theta^U$ is initially close to zero and increases to near $\pi$ before going back close to their initial values. The phases thus do not make a complete circle as $\theta$ from the Wilson loop does at zero temperature. If we raise the temperature further, $\theta^U$ only deviates from its initial value slightly before coming back. Although the winding of $\theta^U$ is obscured by finite-temperature effects, we may still roughly see the non-trivial topology at low $T$. Therefore, the signature of the $Z_2$ index gradually fades away at high enough $T$. As $T\to\infty$, all the density matrix are proportional to the identity matrices. Thus, $V$ is also proportional to the identity matrix and $\theta^U_n=0$ for all $n$.

\subsection{Discussion}
Here we make some approximations to understand the main features of the numerical results. To simplify the calculation, we set $\gamma=0$. We have checked numerically that a small $\gamma$ will not cause qualitative changes. With vanishing $\gamma$, the BHZ model becomes
\be
&&H=R_1\sigma_3\otimes\sigma_1+R_2\sigma_0\otimes\sigma_2+R_3\sigma_0\otimes\sigma_3, \label{eq-BHZ}\\
&&R_1=\sin k_x, \,R_2=\sin k_y,\, R_3=m+\cos k_x+\cos k_y. \nonumber
\ee
Here $\sigma_0$ is the 2 by 2 identity matrix. For later convenience, we denote $\Gamma_1=\sigma_3\otimes\sigma_1$,  $\Gamma_2=\sigma_0\otimes\sigma_2$ and $\Gamma_3=\sigma_0\otimes\sigma_3$.
This model has two doubly degenerate eigenvalues $E=\pm R$ with $R=\sqrt{R_1^2+R_2^2+R_3^2}$. We denote the eigenstates as $\ket{u_1}$ and $\ket{u_2}$ for energy $E=R$ and  $\ket{u_3}$ and $\ket{u_4}$ for $E=-R$. The projectors of the subspace spanned by $\ket{u_{1,2}}$ and $\ket{u_{3,4}}$ are  found to be 
\be
&&P_1=\ket{u_1}\bra{u_1}+\ket{u_2}\bra{u_2}=\frac12(1+\hat{R}_i\Gamma_i), \\
&&P_2=\ket{u_3}\bra{u_3}+\ket{u_4}\bra{u_4}=\frac12(1-\hat{R}_i\Gamma_i),
\ee
where we define $\hat{R}_i=R_i/R$ and the repeated indices imply a summation.

In thermal equilibrium, the distributions of the two subspaces are given by ($k_B=1$)
\be
p_{1,2}=\frac{e^{\mp R/T}}{Z},\qquad Z=4\cosh(R/T).
\ee
From Eq.~(\ref{AU1}), we can obtain the Uhlmann connection by evaluating
\be
A^U_{\mu}=f(R)(P_1 \p_\mu P_2+P_2 \p_\mu P_1)=-\frac{i}2f(R)\epsilon_{ijk}\hat{R}_i\,  \p_\mu\hat{R}_j S_k. \nonumber \\
& &
\ee
Here $\epsilon_{ijk}$ is the Levi-Civita symbol, and the repeated indices imply summation. We also define $S_1=\sigma_0\otimes\sigma_1$, $S_2=\sigma_3\otimes\sigma_2$, $S_3=\sigma_3\otimes\sigma_3$, and  $f(R)=1-\frac{1}{\cosh(R/T)}$.

At the TR invariant momentum points $k_y=0,\,\pi$, we can compute $V$ explicitly because $\hat{R}_2=0$ and find
the vector $\epsilon_{ijk}\hat{R}_i \p_x\hat{R}_j=(0,\,\hat{R_3}\p_x\hat{R}_1-\hat{R_1}\p_x\hat{R}_3,\,0)$ with a fixed direction. Therefore, the path ordering becomes trivial in this case, allowing us to find the Uhlmann-Wilson loop as
\be
V=\exp\Big[-\frac {i S_2}2\int_\mathcal{C} f(R)(\hat{R_3}\p_x\hat{R}_1-\hat{R_1}\p_x\hat{R}_3)d k_x\Big].
\ee
Here $\p_x=\frac{\p}{\p k_x}$, and $\mathcal{C}$ is a contour with fixed $k_y=0,\pi$ while $k_x$ varies from $0$ to $2\pi$. Finally, we obtain
\be
\theta^U=\pm\frac12\int_C f(R)(\hat{R_3}\p_x\hat{R}_1-\hat{R_1}\p_x\hat{R}_3)d k_x.
\ee
At low $T$, $f(R)\approx 1$ and the above integral gives $\pi w$, where $w$ is the winding number of the 2D vector $\vR=(R_1,R_3)$ around the origin $\vR=0$. With the parameters used in the topological case of Figure \ref{W-loop}, we have $|m+\cos k_y|<1$ at $k_y=\pi$. Thus the loop traversed by the tip of $\vR$ encloses $\vR=0$ in this case, giving rise to $w=1$ and $\theta^U\approx\pi$ at $k_y=\pi$. On the other hand, $|m+\cos k_y|>1$ at $k_y=0$, thus the loop of $\vR$ does not enclose $\vR=0$. Therefore, we find $\theta^U\approx0$ at $k_y=0$ as $w=0$. In contrast, the topologically trivial case always has $w=0$. This analysis shows that $\theta^U$ reflects the winding number that characterizes the underlying topology.
At high $T$, $f(R)$ decreases rapidly, causing $\theta_U$ to approach zero. Hence, $\theta^U$ gradually loses its indication of the winding number, as shown in the bottom row of Figure \ref{W-loop}. Further approximate results when $k_y\neq 0,\pi$ can be found in Appendix~\ref{app:UW}, which further catch the features of Figure \ref{W-loop}.

\section{Topology according to Uhlmann phase}\label{sec:Uphase}
The characterization according to the Uhlmann-Wilson loop diminishes as temperature increases, preventing it from being a quantized indicator of the topology at finite temperatures. In the following, we will show that the Uhlmann phase, which is a finite-temperature generalization of the Berry phase, gives quantized values for characterizing the topology at finite temperatures.

\subsection{Topology at zero temperature}
\begin{figure}[t]
\centering
\includegraphics[width=\columnwidth]{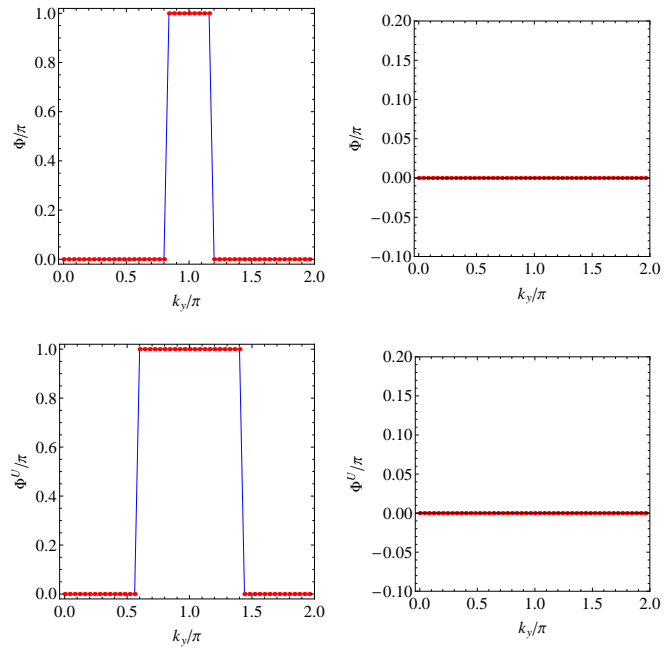}
\caption{Top row: Berry phase $\Phi$ of the BHZ model as a function of $k_y$ with $\gamma=0.2$, $m=0.8$ (left) and $m=2.2$ (right). Bottom row: Uhlmann phase $\Phi^U$ of the BHZ model as a function of $k_y$ with $m=0.8$, $\gamma=0.2$, and $T=0.4$ (left) and $T=2.4$ (right).}
\label{U-loop}
\end{figure}

At zero temperature, the Berry phase may be obtained from the Wilson loop via the expression 
\be
\Phi=\arg\textrm{Tr}\Big[W(k_y)\Big].
\ee
Here $W(k_y)$ is the wilson loop defined in Eq.~(\ref{W-ky}). Since the two occupied states of the BHZ model are almost degenerate, the density matrix is close to the identity matrix. 
From the discussion of section \ref{sec:Wilson}, we know that the eigenvalues of $W(k_y)$ are $\exp[\pm\theta(k_y)]$. Therefore,
\be
\Phi=\arg[\cos\theta(k_y)]=\left\{
                             \begin{array}{ll}
                               \pi, & \pi/2<|\theta(k_y)|<\pi; \\
                               0, & 0<|\theta(k_y)|<\pi/2.
                             \end{array}
                           \right.
\ee
One can see that the Berry phase will jump from zero to $\pi$ when $\theta$ is larger than $\pi/2$. In the top row of Figure \ref{U-loop}, we show the Berry phase $\Phi$ of the BHZ model as a function of $k_y$. For $m=0.8$, there is jump from zero to $\pi$ indicating a non-trivial topological phase. On the other hand, there is no such jump for $m=2.2$, which is topologically trivial. Although the topological regimes from the $Z_2$ index and from the nontrivial Berry phase agree at $T=0$, we would like to mention the subtlety that the Berry phase does not reveal whether $\theta$ from the Wilson loop winds around a full loop or not. In other words, the Berry phase is not another way of expressing the $Z_2$ index. In fact, the Berry phase indicates whether the horizontal lift forms a closed loop~\cite{HouPRA21}, revealing the holonomy group. For the BHZ model, the quantized Berry phase indicates the holonomy forms a $Z_2$ group.

\subsection{Topology at finite temperature}
The Uhlmann phase has been previously used to characterize the topology of two-dimensional Chern insulator at finite temperatures \cite{Viyuela2}. Here we will present the Uhlmann phase of the BHZ model with periodic boundary condition. Suppose we start from a state with amplitude $w_0=\sqrt{\rho_0}$ and parallel transport it along a closed loop to obtain $w_1=\sqrt{\rho_0}V$, where $V$ is from Eq.~(\ref{V-loop}), then the Uhlmann phase is defined as 
\be
\Phi^U=\arg\textrm{Tr}(w_0^\dag w_1)=\arg\textrm{Tr}\Big[\rho_0\,\mathcal{P}\exp(\int_C A^U_\mu d k_\mu)\Big].
\ee
We consider the integration path $C$ from $k_x=0$ to $2\pi$ with fixed $k_y$. The numerical results of $\Phi^U$ for this case is shown in the bottom row of Figure \ref{U-loop} for selected values of temperature. We remark that path ordering appears explicitly in the definition of $\Phi^U$ and needs to be followed in the numerical calculation. 

\begin{figure}
\centering
\includegraphics[width=0.4\textwidth]{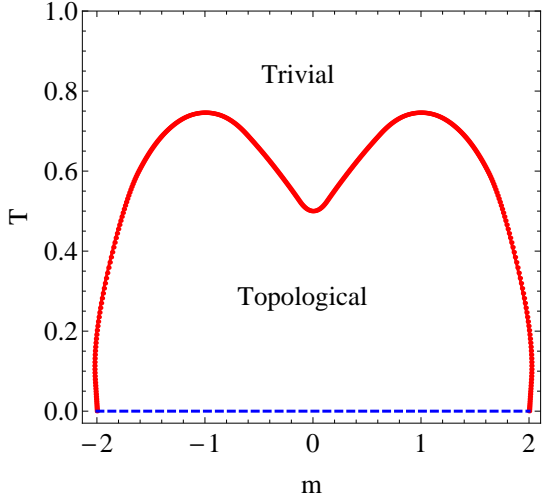}
\caption{Phase diagram of the BHZ model according to the Ulhamnn phase as a function of $m$. Here $\gamma=0.2$. The dashed line at $T=0$ indicates the topological regime according to the $Z_2$ index.}
\label{Tc}
\end{figure}

Importantly, we find that $\Phi^U$ only takes discrete values of $0$ or $\pi$. At low $T$, $\Phi^U=\pi$ inside an interval along the $k_y$ axis. At high $T$, $\Phi^U$ is always zero, regardless of the value of $k_y$. Therefore, we may consider the abrupt jump of $\Phi^U$ from $0$ to $\pi$ as an indicator of the emergence of non-trivial topology in the Uhlmann holonomy, implying a change of the horizontal lift after a cycle~\cite{HouPRA21}. We find that the Uhlmann phase actually provides us a quantized indicator of the topology when compared to $\theta^U$ from the Uhlmann-Wilson loop. Based on $\Phi^U$, we can estimate the transition temperature $T_c$ that separates the trivial and topological regimes as $m$ varies. Here the trivial regime has $\Phi^U=0$ throughout all values of $k_y$ while the topological regime has $\Phi^U=\pi$ in certain range of $k_y$. The phase diagram of the BHZ model is shown in Figure \ref{Tc}. At $T=0$, the topological regime agrees with that from the $Z_2$ index. However, the Uhlmann phase allows a clear distinction between the topological and trivial regimes at finite temperatures. Moreover, we have verified that switching the order of $k_x,k_y$ does not cause any qualitative change to the critical temperature.

\subsection{Discussion}
Again, we will use approximations to understand the Uhlmann phase $\Phi^U$. It can be shown that the density matrix of the BHZ model with $\gamma=0$ in equilibrium is given by 
\be
\rho=p_i P_i=\frac14\Big(1-\tanh(\frac{R}{T})\hat{R}_i\Gamma_i\Big),
\ee
where $\Gamma_i$ with $i=1,2,3$ are the matrices defined below Eq.~(\ref{eq-BHZ}). For convenience, we assume that the initial point of the holonomy corresponds to $k_x=0$. Note that Tr$(\Gamma_i S_j=0)$ for all $i,j$. Combining this with Eq.~(\ref{V-th}), we find
\be
\textrm{Tr}(\rho V)=\cos u,
\ee
which is a real number. Therefore, the Uhlmann phase $\Phi^U=\arg[\textrm{Tr}(\rho V)]$ can only be $0$ or $\pi$ and thus quantized.  Moreover, we have the exact result
\be
\textrm{Tr}(\rho V)=\cos\Big[\frac12\int_C f(R)(\hat{R_3}\p_x\hat{R}_1-\hat{R_1}\p_x\hat{R}_3)d k_x\Big]
\ee
at $k_y=0,\pi$. The Uhlmann phase also indicates the Uhlmann holonomy forms a $Z_2$ group. In the topological case, $u$ can be greater than $\pi/2$ and give rise to a jump of $\Phi^U$ from $0$ to $\pi$.

\begin{figure}
\centering
\includegraphics[width=\columnwidth]{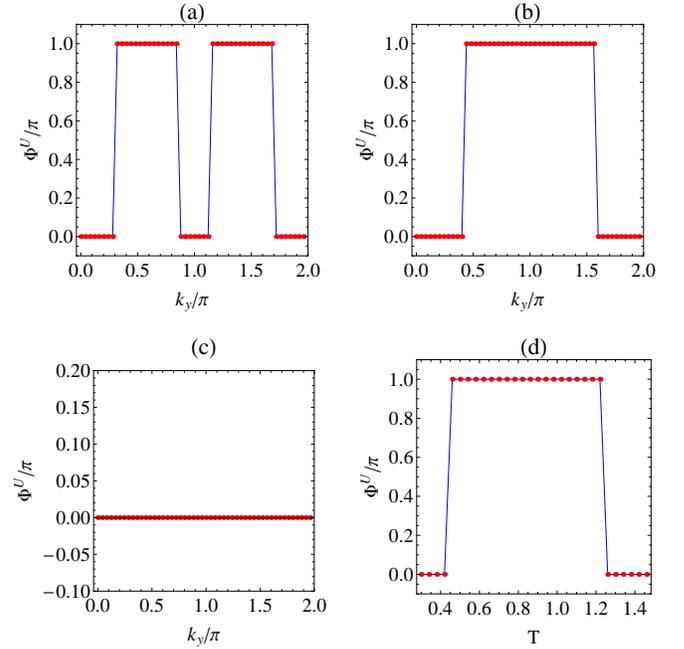}
\caption{Uhlmann phase $\Phi^U$ of the BHZ model with a higher winding number shown in Eq.~(\ref{eq-BHZ-1}) as a function of $k_y$ for panels (a,b,c) and as a function of $T$ for panel (d). For panels (a), (b) and (c), $T=0.4,\,0.8,\,1.4$, respectively. For panel (d), $k_y=\pi$. Here $m=0.8$ and $\gamma=0.2$ for all panels.}
\label{U-2}
\end{figure}

Although both the Berry phase and Uhlmann phase are quantized due to the corresponding holonomy groups, the latter has temperature as an additional tuning parameter. Because of this, a possibility of seeing topologically nontrivial $\Phi^U$ only within some finite-$T$ region emerges. To demonstrate the finite-temperature topological regime, we consider the following generalization of the BHZ model, which introduces a higher winding number:
\be
&&H=R_1\sigma_3\otimes\sigma_1+R_2\sigma_0\otimes\sigma_2+R_3\sigma_0\otimes\sigma_3, \label{eq-BHZ-1}\\
&&R_1=\sin 2k_x, \,R_2=\sin k_y\, R_3=m+\cos 2k_x+\cos k_y.\nonumber
\ee
At $T=0$, the phase $\theta$ of the eigenvalues of the Wilson loop winds around the Brillouin Zone twice. Therefore, the ground state is topologically trivial according to the $Z_2$ index due to the higher winding number. Meanwhile, the Berry phase takes the quantized values of $0$ or $\pi$ as $k_y$ varies. The Uhlmann holonomy varies with temperature and jumps of the Uhlmann phase may occur at finite temperatures. In Figure \ref{U-2}, we plot the Uhlmann phase $\Phi^U$ as a function of $k_y$ of the model at different temperatures and as a function of temperature with fixed $k_y$. One can see that $\Phi^U$ as a function of $k_y$ may jump even number of times at fixed $T$. Importantly, the plot of $\Phi^U$ as a function of $T$ with fixed $k_y$ (in this case, $k_y=\pi$) shows that there exists a topological regime only at finite temperature. Therefore, temperature may introduce topological behavior in systems with higher winding numbers instead of destroying it. Such a finite-temperature topological regime has also been found in a spin-j paramagnet driven by a magnetic field~\cite{Galindo21,HouPRA21}, indicating the generality of temperature-induced topological behavior.

The Uhlmann phase may be observed by constructing the purified states corresponding to the amplitudes of the density matrix in the Uhlmann process~\cite{npj18,HouPRA21}. The purified states are from a composite system consisting of the system of interest and an ancilla. Since the Uhlmann process alone is not compatible with the dynamical evolution according to the system Hamiltonian~\cite{ourPRB20}, one has to impose time evolution operators on both the system and ancilla to render the correct Uhlmann phase for the system alone. On the other hand, interferometric methods \cite{GPMQS1} may be generalized to infer the Uhlmann phase in the future.

\section{conclusion}
\label{sec:Con}
We have presented two finite-temperature generalizations of topological indicators based on the Uhlmann connection and tested them on the BHZ model. The first one extends the Wilson loop at zero-temperature to the Uhlmann-Wilson loop at finite temperatures. While the eigenvalues of the Wilson loop reflects the winding number associated with the $Z_2$ index of the BHZ model, the finite-temperature factor smears out the signature gradually as temperature increases. In contrast, the Uhlmann phase extends the Berry phase to finite temperatures and exhibits quantized values associated with the $Z_2$ group of the Uhlmann holonomy. The quantized values of the Uhlmann phase thus allow us to map out the diagram showing where topological behavior may survive. The comparison of the two indicators shows that finite-temperature quantum systems may exhibit various topological properties characterized by different indicators. Since the two Uhlmann-connection based approaches to finite-temperature topological indicators are general, one can use them to further classify other topological systems away from the zero-temperature limit.

\begin{acknowledgments}
Y. H. was supported by the Natural Science Foundation of China under Grant No. 11874272 and Science Specialty Program of Sichuan University under Grant No. 2020SCUNL210. C. C. C. was supported by the National Science Foundation under Grant No. PHY-2011360.
\end{acknowledgments}

\appendix

\section{Derivation of $A^U_\mu$}\label{app:AU}

For completeness, we reproduce the calculation of $A^U_\mu$ here. As explained in the main text, we consider two density matrices $\rho_1=\rho$ and $\rho_2=\rho+t\p_\mu\rho$. Here $t$ is a small parameter, and those two density matrices are close to each other in the parameter space. Making use of the parallel condition, we find that the following relation:
\be
(U+t\p_\mu U)U^{\dagger}&=&\sqrt{(\rho+t\p_\mu\rho)^{-1}}\sqrt{\rho^{-1}}
\sqrt{\sqrt{\rho}(\rho+t\p_\mu\rho)\sqrt{\rho}}. \nonumber \\
\ee
The Uhlmann connection is then given by
\be
 A^U_\mu &=& \p_\mu U U^{\dagger} \nonumber \\
 &=&\frac{d}{dt}\sqrt{(\rho+t\p_\mu\rho)^{-1}}~\Big|_{t=0}\sqrt{\rho} \nonumber \\
 & &+\rho^{-1}\frac{d}{dt}\sqrt{\sqrt{\rho}(\rho+t\p_\mu\rho)\sqrt{\rho}}~\Big|_{t=0}.
\ee
Now we assume that the eigenvalues and eigenvectors are $E_i$ and $\ket{u_i}$. Using the spectral expansion, the density matrix is given by
\be
\rho=\sum_ip_i\ket{u_i}\bra{u_i},\quad p_i=\frac{e^{-E_i/T}}{\sum_i e^{-E_i/T}}.
\ee
To simplify the expressions, we define $A=\sqrt{(\rho+td\rho)}$ and $B=\sqrt{\sqrt{\rho}(\rho+t\p_\mu\rho)\sqrt{\rho}}$. The following identities can then be derived:
\be
\bra{u_i}\frac{d}{dt}A^2\ket{u_j}\Big|_{t=0}&=&(\sqrt{p_i}+\sqrt{p_j})\bra{u_i}\frac{d}{dt}A\ket{u_j} \nonumber \\
&=&\bra{u_i}d\rho\ket{u_j}, \\
\bra{u_i}\frac{d}{dt}B^2\ket{u_j}\Big|_{t=0}&=&(p_i+p_j)\bra{u_i}\frac{d}{dt}B\ket{u_j} \nonumber \\
&=&\sqrt{p_ip_j}\bra{u_i}\p_\mu\rho\ket{u_j}, \\
\bra{u_i}\frac{d}{dt}A^{-1}\ket{u_j}&=&-\bra{u_i}A^{-1}\frac{dA}{dt}A^{-1}\ket{u_j} \nonumber \\
&=&-\frac{1}{\sqrt{p_ip_j}}\bra{u_i}\frac{d A}{dt}\ket{u_j}.
\ee

Making use of the above identities, we find the matrix elements of $A^U_\mu$ as
\be
&&\bra{u_i}A^U_\mu\ket{u_j}=\sqrt{p_j}\bra{u_i}\frac{d}{dt}A^{-1}\ket{u_j}+p_i^{-1}\bra{u_i}\frac{d}{dt}B\ket{u_j}\nonumber\\
&&=-\frac{1}{\sqrt{p_i}(\sqrt{p_i}+\sqrt{p_j})}\bra{u_i}d\rho\ket{u_j}+\frac{\sqrt{p_ip_j}}{p_i(p_i+p_j)}\bra{u_i}\p_\mu\rho\ket{u_j}\nonumber\\
&&=-\frac{1}{p_i+p_j}\frac{\sqrt{p_i}-\sqrt{p_j}}{\sqrt{p_i}+\sqrt{p_j}}\bra{u_i}\p_\mu\rho\ket{u_j}\nonumber\\
&&=\frac{\bra{u_i}[\p_\mu\sqrt{\rho},\,\sqrt{\rho}]\ket{u_j}}{p_i+p_j}.
\ee
Therefore, we obtained Eq.~(\ref{AU}) in the main text.
The above matrix elements can also be rewritten as
\be
\bra{u_i}A^U_\mu\ket{u_j}&=&-\frac{1}{p_i+p_j}\frac{\sqrt{p_i}-\sqrt{p_j}}{\sqrt{p_i}+\sqrt{p_j}}\bra{u_i}\p_\mu\rho\ket{u_j}\nonumber\\
&=&\frac{1}{p_i+p_j}\frac{\sqrt{p_i}-\sqrt{p_j}}{\sqrt{p_i}+\sqrt{p_j}}(p_i-p_j)\bra{u_i}\p_\mu\ket{u_j}\nonumber\\
&=&\frac{(\sqrt{p_i}-\sqrt{p_j})^2}{p_i+p_j}\bra{u_i}\p_\mu\ket{u_j},
\ee
which give rise to Eq.~(\ref{AU1}).

\section{More topological properties of the BHZ model at $T=0$}\label{app:FK}
The Fu-Kane invariant is constructed from the matrix with elements
\be\label{eq:W}
\bar{W}_{ij}(\vk)=\ep{u_i(-\vk)|U_T K|u_j(\vk)}.
\ee
Here $\ket{u_i(\vk)}$ denotes the eigenstate in momentum space and the indices $i,j$ run through all the occupied bands. $\bar{W}$ is an anti-symmetric matrix at the TR invariant momentum points. The Fu-Kane index is then defined as
\be
I_{FK}=\prod_{i=1}^4\frac{\textrm{Pf}\,[\bar{W}(\vk_i)]}{\sqrt{\det \bar{W}(\vk_i)}}.
\ee
Here the product runs through all the TR invariant momentum points and ``Pf'' denotes the Paffian. One then finds non-trivial or trivial topology corresponding to $I_{FK}=-1$ or $I_{FK}=1$, respectively.

\begin{figure}[t]
\centering
\includegraphics[width=\columnwidth]{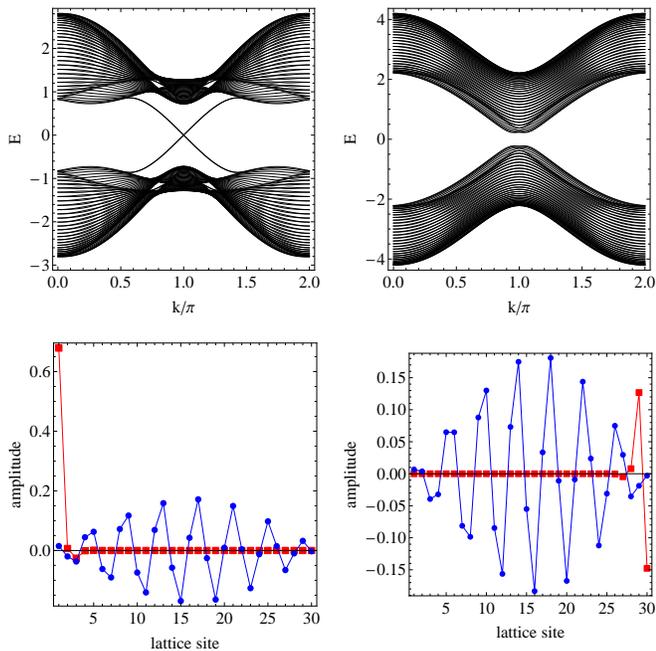}
\caption{Top row: Energy eigenvalues of the BHZ model as a function of $k_y$ with open boundary along the $x$-axis. The left (right) panel corresponds to $m=0.8$ and $\gamma=0.2$ ($m=2.2$ and $\gamma=0.2$) in the topologically non-trivial (trivial) regime. Bottom row: Wave function amplitudes of typical edge (squares) and bulk (circles) states of the spin down component. The right (left) moving edge state is shown in the left (right) panel. Here $k_y=1.2\pi$, $m=0.8$ and $\gamma=0.2$.}
\label{band}
\end{figure}

To see the non-trivial topology of the ground state more clearly, we consider the BHZ model on a cylindrical geometry with open boundary along the $x$-axis and periodic boundary along the $y$-axis. The band structures of the BHZ model are shown in the top row of Figure \ref{band} for a topological case (left panel) and a topologically trivial case (right panel). There are four bulk bands. The bands of spin up and spin down are not exactly degenerate due to the $S_z$ breaking term $H_1$. In the topological case, there exist two curves connecting the valence and conducting bands. Each curve actually corresponds to two degenerate edge states. Thus, there are four edge states in total. To verify the edge states are localized at the open boundaries, we show in the bottom row of Figure \ref{band} the spin down components of two edge states with different propagating directions, which are localized at different edges. This is consistent with the helical-mode behavior, which requires the spin direction to be locked with the velocity direction at a given edge \cite{Qi2}. At finite temperatures, however, all the states are partially occupied due to the Fermi distribution. Therefore, the edge states no longer provide clear indications of topological properties.

\section{Approximate results of Uhlmann-Wilson loop} \label{app:UW}
For $k_y\neq 0,\pi$, we can only find some approximate results of the Uhlmann-Wilson loop $V$. The Uhlmann connection of the BHZ model becomes
\be
A^U_x &=& -i f(R)n_i S_i;\\
n_2&=&\frac12(\hat{R_3}\p_x\hat{R}_1-\hat{R_1}\p_x\hat{R}_3), \nonumber \\
n_1 &=& \frac{R_2\p_x R_3}{2R^2},\quad n_3=-\frac{R_2\p_x R_1}{2R^2}. \nonumber
\ee
For fixed $k_y$, $R_2$ is a constant. Thus, we choose to write the coefficients of $n_i$ in terms of $R_i$ instead of $\hat{R}_i$.
Since all the coefficients of $S_i$ vary with $k_x$, the matrices of $A^U$ at different $k_y$ do not commute with each other. The path ordering becomes a challenge for evaluating the Uhlmann connection. In the case of $k_y\neq 0,\pi$, we can only solve the Uhlmann-Wilson loop from the following differential equation
\be
\frac{d V(k_x)}{d k_x}=A^U_x V(k_x)
\label{dV}
\ee
with the formal solution $V(k_x)=\mathcal{P}\exp\Big(\int_0^{k_x} A^U_x(k'_x) d k'_x\Big)$.
We will focus on the Uhlmann-Wilson loop around a path obtained by $V=V(k_x=2\pi)$ as $k_x$ goes from $0$ to $2\pi$.

For fixed $k_y$, $R$ only weakly depends on $k_x$. Thus, we would expect that the trajectory of the vector $(n_1,n_3)\propto(\sin k_x,\cos  k_x)$ as a function of $k_x$ is roughly like a circle.
We can make a gauge transform to convert the above circle to a single point. Suppose $V= U(k_x)V'$, then the equation becomes
\be
\frac{d V'}{d k_x}=\Big(U^{\dag}A_U U-U^{\dag}\frac{d U}{d k_x}\Big)V'.
\ee
Next, we rewrite $A^U_x$ as
\be
A^U_x=-i f(R)\Big[n_2S_2
-\frac{R_2}{2R^2}e^{-ik_x S_2/2}S_3e^{ik_x S_2/2}\Big].
\ee
Now it is clear that we can choose $U=e^{-ik_x S_2/2}$ to find
\be
\frac{d V'}{d k_x}=\Big\{-i f(R)\Big[n_2S_2
-\frac{R_2}{2R^2}S_3\Big]+\frac{i}2 S_2\Big\}V'.
\ee
Comparing with the $k_y=0,\pi$ cases, there are two extra terms in the transformed $A^U_x$ with almost constant components. If we approximate the trajectory of $\hat{R}_i$ by a constant-latitude circle on a unit sphere, then the vector $n_2$ is also almost constant. With the above consideration, we can make a crude approximation to solve the equation by ignoring the non-commuting parts of $A^U$ at different values of $k_x$ and obtain the final results. Explicitly,
\be
V&\approx&\exp\Big[i(-u_2S_2+u_3S_3)\Big], \label{V-th}\\
u_2&=&\frac12\int_C f(R)(\hat{R_3}\p_x\hat{R}_1-\hat{R_1}\p_x\hat{R}_3)d k_x,\nonumber \\
u_3&=&\sin k_y\int_C \frac{f(R)}{2R^2}d k_x, \nonumber
\ee
which can also be expanded as
\be
V\approx\cos u+i(-\hat{u}_2S_2+\hat{u}_3S_3)\sin u.
\ee
Here $\hat{u_i}=u_i/u$ and $u=\sqrt{u_2^2+u_3^2}$. It can be shown that the eigenvalues of $V$ are $e^{\pm i u}$ with double degeneracy. Then, the arguments of the eigenvalues are $\theta^U=\pm u$ and satisfy the relation of Eq.~(\ref{th-U}). We would expect that $\theta^U$ will interpolate the results between $\theta^U\approx0$ at $k_y=0$ and $\theta^U\approx\pi$ at $k_y=\pi$. The correction due to $u_3$ is negligible around $k_y=0,\pi$ but becomes largest around $k_y=\pi/2$. Thus, the above discussion roughly explains the qualitative behavior of $\theta^U$ as a function of $k_y$.

\bibliographystyle{apsrev}

\end{document}